\documentstyle[11pt,aaspp4,flushrt]{article}

\input{epsf}

\slugcomment{Version of July 20, 1999}

\lefthead{Gaite, Dom\'\i nguez, P\'erez-Mercader} 
\righthead{Transition to homogeneity in the galaxy distribution}

\begin{document}

\title{The fractal distribution of galaxies and the transition to
  homogeneity}

\author{Jos{\'e} Gaite, Alvaro Dom\'\i nguez, and Juan
  P\'erez-Mercader} \affil{Laboratorio de Astrof\'\i sica Espacial y
  F\'\i sica Fundamental (INTA-CSIC), Apartado 50727, 28080 Madrid}
\affil{gaite@laeff.esa.es, dominguez@laeff.esa.es,
  mercader@laeff.esa.es}


\begin{abstract}
  There is an ongoing debate in cosmology about the value of the
  length scale at which ``homogeneity'' in the matter distribution is
  reached or even if such a scale exists. In the wake of this debate,
  we intend in this letter to clarify the meaning of the statement
  {\it transition to homogeneity} and of the concept of {\it
    correlation length}. We show that there are two scales each
  associated to the fractal and to the homogeneous regimes of the
  matter distribution, respectively. The distinction between both
  scales has deep consequences, for there can be a regime which
  despite having small fluctuations around the average density
  exhibits large clusters of galaxies.
\end{abstract}

\keywords{methods: statistical --- galaxies: clusters: general ---
  large-scale structure of universe}

\section{Introduction}

The analysis of the distribution of large-scale structure in the
Universe is done by applying statistical tools. The most basic
descriptor is, aside from the average mass density, $\bar \rho =
\langle \rho ({\bf r}) \rangle$, the reduced two-point correlation
function (\cite{Peebook}),
\begin{equation}
\xi(r)=\frac{\langle \rho({\bf r})\,\rho({\bf 0})\rangle}{\bar \rho ^2}-1 
= \frac{\langle \delta \rho ({\bf r})\,\delta \rho ({\bf 0})\rangle}
{\bar \rho ^2},  
\end{equation}
where we assume statistical homogeneity and isotropy. Higher order
correlation functions of the probability distribution are also of
importance, but their determination becomes less reliable as the order
increases.  However, for the purpose of illustrating our point, it
will suffice to concentrate on $\xi(r)$. The study of galaxy
catalogues leads to fitting $\xi(r)$ by a power law on the length scales
probed by these catalogues,
\begin{equation}
\xi(r) = \left({r_0 \over r}\right)^{\gamma} ,
\label{observedxi}
\end{equation}
but the values of $r_0$ and $\gamma$, and even whether the latter is
scale-dependent, are still being debated (\cite{Pietro}; \cite{Sylos};
\cite{Davis}; \cite{Guzzo}; \cite{Scara}; \cite{Martinez}). At any
rate, it seems that scale-invariance is approximately true on a
non-negligible range of scales.

The distance $r_0$, defined by $\xi(r_0)= 1$ (or by some other
qualitatively equivalent condition, e.g.\ \cite{Martinez}), has been
called ``correlation length'' in the literature. However, it is well
known in the theory of critical phenomena (\cite{Stanley};
\cite{Dowetal}; \cite{Ma}) that power-law correlated fluctuations let
themselves be felt over the entire range of observation. There, the
physical meaning of a correlation length corresponds to the length
scale where power-law decay crosses over to exponential decay, and has
nothing to do with the condition $\xi \sim 1$.

The distance $r_0$ has certainly the meaning of separating a regime of
large fluctuations, $\delta\rho \gg {\bar \rho}$, from a regime of
small fluctuations, $\delta\rho \ll {\bar \rho}$. The regime of small
fluctuations has been identified with a homogeneous distribution and,
because of this, $r_0$ has been called the scale of transition to
homogeneity. However, as we will see, the sense in which $\delta\rho
\ll {\bar \rho}$ leads to homogeneity is not equivalent to absence of
structure. In fact, fluctuations correlated by a power law, albeit
small, may give rise to collective, macroscopic phenomena, such as for
instance critical opalescence in fluids (\cite{Stanley};
\cite{Dowetal}). {\it Absence of structure is not determined just by
  the condition that correlations vanish at large distances, but
  rather by how fast they do so.} This is measured by the correlation
length $\lambda_0$, separating slow from fast decay. Fast decay is
generally identified with exponential decay, but it could also refer
to another type of sufficiently rapid decay.

The point, therefore, is to establish a clear distinction between the
{\em strength} of the correlations (a measure of how {\em large} the
fluctuations around the mean can be), and the {\em range} of the
correlations (a measure of how {\em far} correlations are felt). These
two characteristics of the correlation function, though related, are
not equivalent to each other. This subtle difference has been usually
overlooked in the cosmological literature or, at least, not clearly
stated.

\section{Relevant concepts}

In order to exemplify the main concepts, we consider a random
distribution of particles whose average density is $\bar \rho$. In the
next section we will discuss the implications for cosmology. The
second moment of the density field is given by
\begin{equation} 
G(r) = \bar \rho ^2\, [1 + \xi(r)] .
\label{2moment}
\end{equation}
Correlations are weak if $\xi<1$, and strong if $\xi>1$. Thus the
scale $r_0$ defined by $\xi(r_0)=1$, which separates these two
regimes, is related to the strength of the correlations. In the
parlance of structure formation, $r_0$ is the scale separating at a
given time the linear from the nonlinear dynamical regimes
(\cite{SahCo}), and we may call it the {\em scale of nonlinearity}. A
more intuitive picture can be achieved if one examines the variance of
the smoothed density field, $\rho_L$, the number of particles
contained in a volume of size $L^3$ per unit volume. This variance is
computed as an integral of $\xi(r)$. We consider the cosmologically
interesting case where the correlation function has a simple
scale-invariant form like (\ref{observedxi}), with given $r_0$ and
$\gamma<3$ (this latter requirement assures convergence of the
integral). Then we can write:
\begin{equation} 
{\langle (\delta \rho_{L})^2 \rangle \over \bar \rho^2} \sim \left(
{a \over L} \right)^3 + B \left({r_0 \over L}\right)^{\gamma} , 
\label{relflucL}
\end{equation} 
where we have defined a length scale $a=\bar \rho ^{-1/3}$ ($a^3$ is
the volume per particle), and $B$ is a positive numerical constant of
order one. The first term is the shot-noise contribution, and the
second is due to correlations. We conclude that if the volume
containing the sample is large enough ($L \gg r_0, a$), the relative
fluctuations of $\rho_L$ around its mean are very small, and a good
estimate of $\bar \rho $ can be extracted from the sample. What
happens for {\em intermediate} values of $L$ depends on the {\em
  relative} values of $a$ and $r_0$. We consider two physically
interesting cases: (i) $a \ll r_0$, as may be the case with the galaxy
distribution, so that for scales $a \ll L \lesssim r_0$ fluctuations
are large and the correlation term dominates; and (ii) $a \sim r_0$,
as in a fluid in thermal equilibrium, so that for scales $L \lesssim
a, r_0$, fluctuations are large but dominated by shot noise. In
either case no reliable estimate of $\bar \rho$ can be extracted. We
also see that a Gaussian approximation to the one-point probability
distribution of $\rho_L$ is certainly excluded in these cases, because
$\rho_L \geq 0$ by definition and hence the probability distribution
becomes prominently skewed if $\delta \rho_L \gtrsim \bar \rho$.

On the other hand, the range of the correlations is related to how
fast they decay at infinity, which is measured by the spatial moments
of the two-point correlation function:
\begin{equation}
  {\cal M}_n \equiv  \int d {\bf r} \, {\bf r} \stackrel{(n)}{\cdots} 
  {\bf r} \, \xi(r) \propto \lim _{k \rightarrow 0} \nabla_{\bf k} 
  \stackrel{(n)}{\cdots} \nabla_{\bf k} P(k) , 
\end{equation}
where we have introduced the Fourier transform of the two-point
correlation function, the power spectrum $P(k) = \int d{\bf r} \, e^{i
  {\bf k} \cdot {\bf r}} \,\xi(r)$. Notice that ${\cal M}_n$ vanishes
for odd $n$ due to isotropy of the correlations. Starting from the
interpretation of $\xi(r)$ as the density contrast of particles at a
distance $r$ from any one particle, it is straightforward to give a
physical meaning to the moments ${\cal M}_n$: they are the multipoles
characterizing the morphology of the typical cluster generated by
correlations. Correlations are short-ranged if all the moments are
finite, which implies that they decay faster than any power of the
distance and that $P(k)$ is {\em analytic} at ${\bf k} = {\bf 0}$. One
can then define a typical size of the cluster, $\lambda_0$, the {\em
  correlation length}, as
\begin{equation}
  \lambda_0^2 = {1 \over 2} \left| {Tr \, {\cal M}_2 \over {\cal M}_0} 
  \right| = \lim _{k \rightarrow 0} {1 \over 2} 
  \left| {\nabla^2_{\bf k} P(k) \over P(k)} \right| ,
\end{equation}
in agreement with the standard definition in condensed matter physics
(see e.g.\ \cite{Ma}); if $P(0)$ or $\nabla^2_{\bf k}P(0)$ vanishes, then the
definition of $\lambda_0$ must be generalized in terms of a different
quotient between higher-order, non-vanishing moments. Therefore, if
one observes a realization of the field $\varrho_L$ with a smoothing
length $L \gg \lambda_0$, then clusters are unobservable and the
distribution is indistinguishable from the case that there are no
correlations at all. Conversely, if $\xi(r)$ decays as a power of $r$,
so that it is long-ranged, then the moments ${\cal M}_n$ diverge from
some $n$ onwards (depending on the exponent of decay), and $P(k)$ is
not analytic at ${\bf k} = {\bf 0}$. Hence, no matter how large $L$
is, the existence of clusters will always be detectable, so that one
associates an infinite correlation length to a power-law decay. In
conclusion, {\it it is $\lambda_0$ the scale that marks the transition
  to homogeneity in the sense of absence of structure.}  Notice that
in principle $\lambda_0$ and $r_0$ are different and in fact they can
even be orders-of-magnitude different, as a power-law decay of
$\xi(r)$ exemplifies.

It is clear now that there are in general at least three different
length scales ($a$, $r_0$, $\lambda_0$), thus defining four regimes,
which we now discuss. Notice that without further knowledge of the
physics of the system, one cannot exclude that any two of the scales
are of the same order, which is not a rare case in physics, so that
some of the regimes below can be difficult to observe. In what follows
we will assume the case most appropriate in a cosmological context,
that $a < r_0 < \lambda_0$.  We then have:
\begin{itemize}
\item {\bf Homogeneous regime}, for scales $r \gg \lambda_0$. On these
  scales, fluctuations are small ($\xi(r) \ll 1$) and the system
  structureless. Typically, correlations are exponentially damped and
  $\rho_L$ has a prominently peaked distribution, so that the mean
  density $\bar \rho$ can be reliably estimated. An example is a
  liquid in thermal equilibrium far from a phase transition, for which
  $\xi(r) \sim (r_0/r) \exp(-r/\lambda_0)$, where $a$, $r_0$ and
  $\lambda_0$ are all of the order of the molecular diameter
  (\cite{Landau}; \cite{Goodstein}).
\item {\bf Critical regime}, for scales $r_0 \ll r \ll \lambda_0$.
  Fluctuations are still small, so that the mean density $\bar \rho$
  can be estimated with confidence too but now because fluctuations
  are correlated and therefore behave coherently on these scales, they
  give rise to spatially extended structures. Moreover, if the
  correlations follow a power law on these scales, fluctuations about
  the background have a self-similar, fractal structure. An example is
  a fluid at the critical point, for which $\xi(r)$ is like in
  equation~(\ref{observedxi}) with $\gamma < 3$. Again $a$ and $r_0$
  are of the order of the molecular diameter, but now $\lambda_0
  \rightarrow \infty$. Notice that small fluctuations do not imply
  uninteresting behavior: since they are macroscopic in spatial
  extent, one observes phenomena such as critical opalescence
  (\cite{Stanley}; \cite{Dowetal}). The reader may feel uneasy with
  the apparent contradiction between the claim that fluctuations are
  small in a critical fluid and the standard claim that they are in
  fact large.  The point is to identify with which quantity one is
  comparing the fluctuations. When compared with $\bar \rho $, as we
  do, they are small, which is a prerequisite to render a
  thermodynamic description valid. When compared with typical
  fluctuations far from the critical point, they are extremely large,
  and in fact the ratio (size of critical fluctuations)/(size of
  non-critical fluctuations) usually diverges in the thermodynamic
  limit, as is seen from equation~(\ref{relflucL}).
\item {\bf Fractal regime}, for scales $a \ll r \ll r_0$. Now
  fluctuations exhibit structure and at the same time are very large
  ($\xi \gg 1$), so that from equation~(\ref{2moment}) it follows
  that $G(r) \approx \bar \rho ^2 \xi(r)$, and $\bar \rho $ {\em
    cannot} be estimated. For power-law correlations, this means that
  it is the density itself (and not only its fluctuations around the
  mean) that appears to have fractal structure (or multifractal,
  depending on how higher-order correlations scale). This manifests
  itself in that most realizations of the stochastic field $\rho({\bf
    r})$ do not look like a continuum field at all on these scales,
  but are very irregular and full of voids of every size separated by
  very high density condensations. This regime is not encountered in a
  critical fluid because $r_0$ is of the order of the molecular
  diameter, but may be relevant in cosmology, as we will see.
\item {\bf Shot-noise regime}, for scales $r \ll a$. The presence of
  correlations is masked by shot noise: fluctuations are very large,
  but simply because the discrete nature of the underlying
  point-process is dominant.
\end{itemize}

It may be helpful at this point to show pictorially a distribution in
the critical regime. Figure~1 is a snapshot of a critical
two-dimensional lattice gas,\footnote{It actually is isomorphic to the
  two-dimensional Ising model. The three-dimensional Ising model as a
  description of the galaxy distribution has been introduced in
  \cite{DavJuan}; \cite{PGHL}.} with $\lambda_0 \rightarrow \infty$,
but $a$ and $r_0$ are both of the order of a pixel (=atom), barely
discernible to the naked eye. In a critical fluid the average density
is well defined, but one can clearly see in the figure structures
(i.e., inhomogeneities) on scales much larger than $r_0$. Although
one may observe what seem like voids of arbitrary size, a closer
examination unveils that the voids contain clusters of atoms. A
fractal however must contain voids of every size, owing to scale
invariance, and the voids need to be {\em totally} empty (density
exactly zero in them), so the fluctuations of the density over a
volume of size $L$ are large at any scale. It is now clear that what
has scale invariance, or fractal character, in a critical fluid is
{\it the fluctuations over the average density}, rather than the
density field itself. For comparison, Figure~2 shows the same
model at very high temperature, with $\lambda_0$ of the order of a
pixel (comparable to $a$ and $r_0$), and where it exhibits Poissonian
fluctuations. Visual inspection shows immediately that fluctuations
are less prominent than in the critical regime, and that there is no
structure at all.

\placefigure{fig1}

\placefigure{fig2}

\section{Implications for cosmology}

We now apply the above to cosmology. We have two {\em different}
scales, $r_0$ and $\lambda_0$, related to the intensity and to the
range of the correlations, respectively. This does not preclude the
existence of other intermediate length scales, which do not however
affect the physical meaning of $r_0$ and $\lambda_0$, nor the
discussion to follow. Incidentally, one of these scales, which has
also been termed correlation length, is defined as (\cite{KPSM};
\cite{SSMPM})
\begin{equation}
  R_{\phi}^2 = 6\, \frac{\langle \phi^2 \rangle}{\langle (\nabla \phi)^2 
    \rangle} = 6 \,\frac{\int d{\bf k} \; k^{-4} P(k)}{\int d{\bf k} 
    \; k^{-2} P(k)} ,
\end{equation}
where $\phi$ is the peculiar gravitational potential due to the
fluctuations in the density field, $\delta \rho$. This length can be
understood as the characteristic scale of spatial variation of the
field $\phi$. But this is not the correlation length either, because a
correlation between the values of the field at two separate points
does not imply in general that these values must be similar.

As discussed in the previous section, the length scale $\lambda_0$ is
given by the behavior of the power spectrum on the largest scales.
Since one cannot observationally gather information about scales
larger than the horizon, $\lambda_0$ can be estimated only by {\em
  assuming} that the known behavior of $P(k)$ on the largest
observable scales holds beyond the horizon. The power spectrum on the
largest scales can be extracted from the analysis of the microwave
background, yielding $P(k) \sim k^{1.2}$ (\cite{COBE}; \cite{Smoot})
at the epoch of decoupling. If this non-analytic behavior were
extrapolated to $k \rightarrow 0$, it would imply $\lambda_0 =
\infty$. The physical interpretation of this conclusion is that the
correlation length was certainly larger than the horizon at the epoch
of decoupling. Hence, since causal processes can create correlations
up to the scale of the horizon only, we must be observing the
remaining, frozen imprints of a pre-inflationary epoch which have not
been erased yet. Moreover, given that the evolution of the power
spectrum by gravitational instability does not alter its shape in the
linear regime (\cite{Peebook}), one must conclude that $\lambda_0$
{\em is} still beyond the horizon in the present epoch.  Thus, it may
seem that the {\em homogeneous regime} is beyond reach and that the
largest observable scales in the Universe belong to the {\em critical
  regime}.  As a matter of fact, notice that the well-known pictures
of the microwave background by COBE (\cite{COBE}) resemble figure~1
much more than figure~2.

Let us consider now the scale of nonlinearity, $r_0$. The smallness of
density fluctuations as inferred from observations of the microwave
background ( $\langle (\delta \rho_{L})^2 \rangle / \bar \rho_{L}^2
\sim 10^{-7}$ for $L \sim 10^3 \, h^{-1}$ Mpc) yields an upper bound
on $r_0$. Smaller scales can be probed by means of the galaxy
catalogues. The conclusions related to the matter distribution depend
on the bias parameter, but, in agreement with theoretical models, this
parameter can be assumed to be of order unity. This implies in turn
that $r_0$ can be directly extracted from the galaxy-galaxy
correlations. The standard result therefore is that $r_0 \approx 5\,
h^{-1}$ Mpc (\cite{DavPeeb}; \cite{Peebook}), so that currently
available galaxy catalogues, of size $L \gg r_0$, are probing the {\em
  critical regime}. Hence, the galaxy distribution should look very
much like a fluid at the critical point: small but spatially extended
fluctuations. As already stressed, this does not mean that the
catalogues are probing the transition to homogeneity, since $r_0$ is
not the correlation length.

There has recently arisen, however, a controversy on the validity of
$r_0 \approx 5 \, h^{-1}$ Mpc, due to suggestions that $r_0$ is in
fact larger than the maximum effective size of the current largest
galaxy catalogues, and that a simple power-law form like
(\ref{observedxi}) with a fixed exponent holds at least up to this
maximum scale (\cite{Pietro}; \cite{Sylos}). If so, then the
catalogues are probing the {\em fractal regime}. As shown in the
previous section, this implies that the mean galaxy number density
cannot be reliably estimated from these catalogues, and hence neither
can $r_0$. This means that they do not allow to discriminate between a
distribution that has a nonvanishing $\bar \rho$ and {\em looks}
fractal up to scales of the order of $r_0$, and a distribution whose
$\bar \rho$ vanishes and thus {\em is} fractal up to arbitrarily large
scales. Not surprisingly, the idea of a universe with a vanishing
average density as the size of the sample volume increases is very
old, dating back to at least 1908 (\cite{Charl}). This construction
was based on a hierarchy of clustering without limit, that is,
clusters of galaxies which form clusters of clusters, which in turn
also form clusters and so on, {\em ad infinitum}. It is illustrative
to read now the old review by G. de Vaucouleurs advocating for a
hierarchical cosmology (\cite{Vau}).

As we have seen before, it appears that the correlation length
$\lambda_0$ is much larger than the scale of nonlinearity, $r_0$. This
could lead to interesting consequences. As already remarked, and
illustrated by the discussion of figures~1 and 2,
there can exist structure in the critical regime, i.e. on intermediate
scales $r_0 \ll r \ll \lambda_0$. Hence, this could be the reason why
clustering is observed, as galaxy clusters and superclusters, even on
scales much larger than $r_0$, which is a ``paradox'' in the
interpretation of galaxy catalogues. Similarly, the increase of $r_0$
with bias threshold should not be interpreted in terms of a larger
correlation length for the corresponding structure (\cite{pir}).

Another consequence of a large $\lambda_0$ is related to the effect of
gravitational lensing on the microwave background (\cite{BlSc};
\cite{Kash}; \cite{CoEf}; \cite{MSS}; \cite{FME}; \cite{MSC}). The
geodesic equation for light propagation is formally identical to the
geometric-optics equation for light-rays in a medium of non-uniform
refractive index, which allows one to identify the gravitational
potential with an effective refractive index. Therefore, the fact that
the correlation length of the potential diverges means that
gravitational lensing of the microwave background could exhibit
an effect akin to critical opalescence (a sort of ``cosmological
opalescence'', \cite{unpub}).

In conclusion, we have provided a clear explanation and a precise
mathematical formulation of the difference between the intensity of
fluctuations and their spatial extent in the context of
cosmological structures. These two properties can be characterized by
two different length scales: the scale of nonlinearity, $r_0$, and the
correlation length, $\lambda_0$. We have discussed the behavior of the
correlations in the different regimes which these two lengths define
and the connection with cosmological observations. It should be clear
now that the statement ``transition to homogeneity'' as used in the
cosmological literature really means ``transition to the regime of
small-amplitude fluctuations'', which {\em does not} imply absence of
structure. In fact, we have also argued that this remaining ``critical
structure'' could have non-trivial consequences for the interpretation
of observations. For instance, no real transition to homogeneity is
likely to be observed in the galaxy distribution.

\acknowledgements

We would like to acknowledge useful discussions with M. Kerscher and
conversations with M. Carri\'on, R. Garc\'{\i}a-Pelayo, J.R. Acarreta,
J.M. Mart\'{\i}n-Garc\'{\i}a and L.  Toffolatti, as well as remarks on
the manuscript by T. Buchert, and correspondence with L. Guzzo and E.
Gawiser. We also thank the anonymous referees for ``critical''
discussions on the manuscript. J.G. acknowledges support under Grant
No.\ PB96-0887.

\newpage
\epsfxsize=8.5cm
\epsfbox{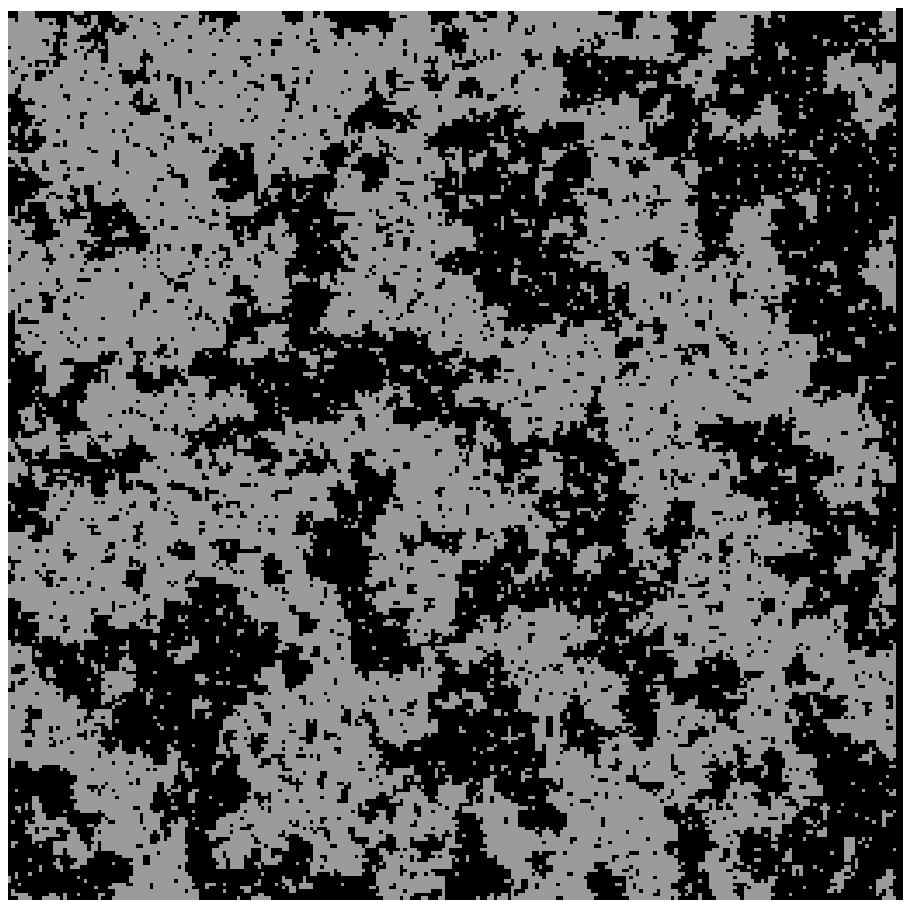}
Fig.\ 1: Critical fluid.

\vspace{12mm}
\epsfxsize=8.5cm
\epsfbox{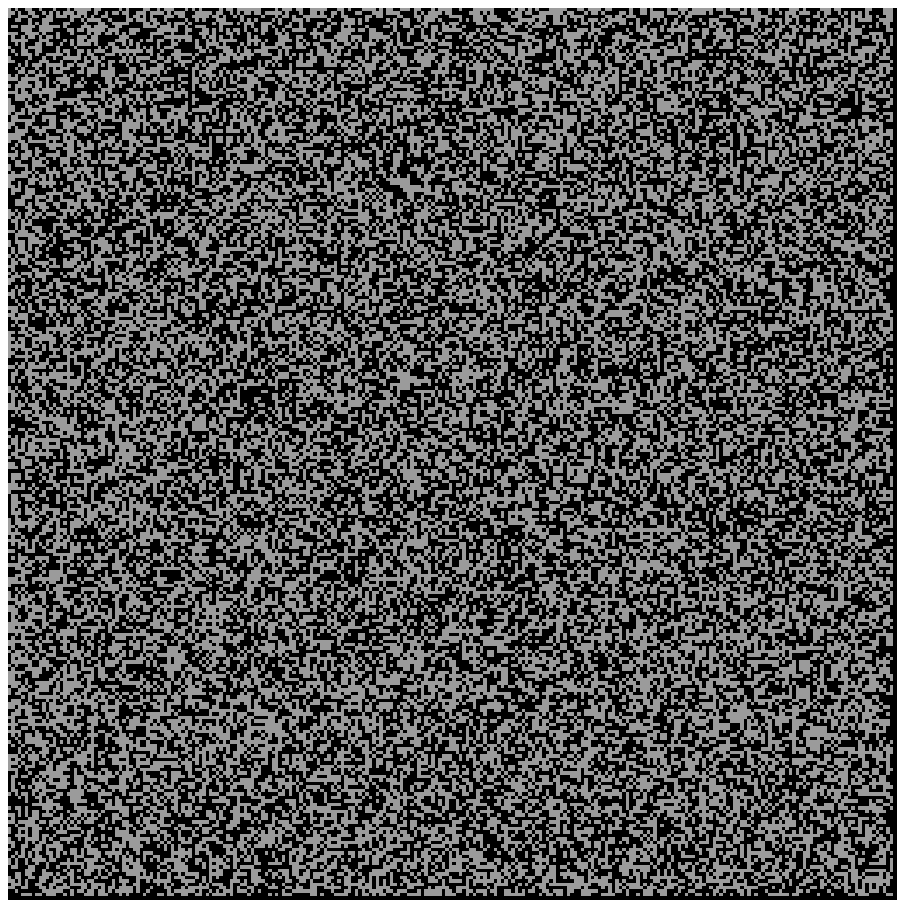}

Fig.\ 2: Non-critical fluid.

\end{document}